\documentclass[runningheads,a4paper]{llncs}

\usepackage{array} 
\usepackage{times}
\usepackage{url}
\usepackage{latexsym}
\usepackage{times}
\usepackage{latexsym}
\usepackage{amsmath}
\usepackage{url}
\usepackage[mathscr]{euscript}
\usepackage{multirow,tabu}
\usepackage{arydshln}
\usepackage{color, colortbl}
\usepackage{graphicx}
\usepackage{caption}
\usepackage{subcaption}
\usepackage[margin=1.68in]{geometry}
\usepackage[dvipsnames,table]{xcolor}
\usepackage{amssymb}
\usepackage{tabularx}
\usepackage{makecell}
\usepackage{hyperref}

\newcommand{\jComment}[1]{\textcolor{black}{#1}} 

\newcommand{\specialcell}[2][c]{\begin{tabular}[#1]{@{}c@{}}#2\end{tabular}}

\usepackage{url}
\urldef{\mailsa}\path|{alfred.hofmann, ursula.barth, ingrid.haas, frank.holzwarth,|
\urldef{\mailsb}\path|anna.kramer, leonie.kunz, christine.reiss, nicole.sator,|
\urldef{\mailsc}\path|erika.siebert-cole, peter.strasser, lncs}@springer.com|

\begin{document}

\mainmatter  

\title{Dimension Projection among Languages based on Pseudo-relevant Documents for Query Translation}

\titlerunning{Dimension Projection among Languages for Query Translation}

%
%
\author{Javid Dadashkarimi%
	\and  Mahsa S. Shahshahani \and Amirhossein Tebbifakhr \and Heshaam Faili \and Azadeh Shakery}

\urldef{\mailsa}\path|{dadashkarimi,ms.shahshahani,a.tebbifakhr,hfaili,shakery}@ut.ac.ir|   

\author{Javid Dadashkarimi, Mahsa S. Shahshahani, Amirhossein Tebbifakhr, Heshaam Faili, and Azadeh Shakery\\
School of ECE, College of Engineering, University of Tehran, Iran\\
\mailsa}

%
\authorrunning{Dadashkarimi et al.}


%
%

\toctitle{Lecture Notes in Computer Science}
\tocauthor{Authors' Instructions}
\maketitle

\begin{abstract}
Using top-ranked documents in response to a query has been shown to be an effective approach to improve the quality of query translation in dictionary-based cross-language information retrieval. 
In this paper, we propose a new method for dictionary-based query translation based on dimension projection of embedded vectors from the pseudo-relevant documents in the source language to their equivalents in the target language.
To this end, first we learn low-dimensional vectors of the words in the pseudo-relevant collections separately and then aim to find a query-dependent transformation matrix between the vectors of translation pairs appeared in the collections.
At the next step, representation of each query term is projected to the target language and then, after using a softmax function, a query-dependent translation model is built. 
Finally, the model is used for query translation.
Our experiments on four CLEF collections in French, Spanish, German, and Italian demonstrate that the proposed method outperforms a word embedding baseline based on bilingual shuffling and a further number of competitive baselines.
The proposed method reaches up to 87\% performance of machine translation (MT) in short queries and considerable improvements in verbose queries.
\end{abstract}

\section{Introduction}
\vspace{-0.2cm}
Pseudo-relevance feedback (PRF) has long been shown to be an effective approach for updating query language models in information retrieval (IR)~\cite{Lv:2014,Zamani:2016,Lavrenko:2001,Lavrenko:2002}. 
In cross-language environments \jComment{where} there are a couple of document sets in different languages, building such models requires bridging the gap of the languages. To this end, cross-lingual topical relevance models (CLTRLM) aims to find a way to transform knowledge of the sets to the query model using bilingual topic modeling and a bilingual dictionary~\cite{Lavrenko:2002,Ganguly:2012}.
The performance of CLTRLM is heavily depends on the number of alignments in a comparable corpora and their qualities as well.
Recently, bilingual word embedding is tailored effectively to this end where low-dimensional vectors are built after shuffling all the alignments \cite{Vulic:2015b}.
However the effectiveness of this method has not been investigated at cross-lingaul PRF yet.

In this paper we propose a new method for building translation models on pseudo-relevant collections using a neural network-based language model. 
The proposed cross-lingual word embedding translation model (CLWETM) takes advantage of a query-dependent transformation matrix between low-dimensional vectors of the languages.
Indeed, we aim to find a transformation matrix to bring the vector of each query term, built on the source collection, to dimensionality of the target language and then compute the translation probabilities based on a softmax function.
To this aim, first we learn word representations of the pseudo-relevant collections separately and then focus on finding a transformation matrix minimizing a distance function between all translation pairs appeared at the collections.
This method captures semantics of both the collections with a rotation and a scaling embedded in the matrix.
Finally, a softmax function is used to build a query-dependent translation model based on similarity of transformed vector of each query term with \jComment{the vectors of its} translation candidates in the target language.

Unlike CLTRLM and the mixed word embedding translation model (MIXWETM) based on shuffling alignments in a comparable corpora (\cite{Vulic:2015b}), CLWETM considers quite sentence-level contexts of the words and therefore  captures deeper levels of n-grams in both the languages.
Furthermore, the obtained model can be incorporated within a language modeling framework, the state-of-the-art retrieval framework in the literature, and therefore the proposed method does not suffer from disadvantages of low-dimensional query-document similarity in ad-hoc retrieval~\cite{Vulic:2015b}.

Experimental results on four CLEF collections in French, German, Spanish, and Italian demonstrate that the proposed method outperforms all competitive baselines of dictionary-based cross-language information retrieval (CLIR) in language modeling when it is combined with a global translation model.
The proposed method reaches up to 83\% performance of the monolingual run and 87\% performance of machine translation in short queries.
CLWETM has more better results in verbose queries and even improvements compared to MT in the Italian collection.

\vspace{-0.4cm }
\section{Related Works}
\vspace{-0.2cm }
\label{sec:related works}
Pseudo-relevance feedback has long been employed as a powerful method for estimating query language models in a large number of studies~\cite{Lv:2014,Lv:2010,Zamani:2016}.
Cross-lingual relevance model (CLRLM) and CLTRLM are state-of-the-art methods in cross-lingual environments~\cite{Ganguly:2012,Vulic:2015a,Lavrenko:2002}. 
Unlike CLRLM that depends on parallel corpora and bilingual lexicons, CLTRLM aims at finding a number of bilingual topical variables from a comparable corpus in order to transfer relevance score of a term from one language to another.
Ganguly et al. proposed to use these variables for query translation and demonstrated that CLTRLM is an effective method particularly for resource-lean languages~\cite{Ganguly:2012}. 

In CLTRLM top-ranked documents $F^s=\{d^s_1,d^s_2,..,d^s_{|F^s|}\}$ retrieved in response to the source query ($\textbf{q}^s$) and top-ranked documents $F^t=\{d^t_1,d^t_2,..,d^t_{|F^t|}\}$ retrieved in response to a translation of the query ($\textbf{q}^t$) are assumed to be relevant documents and then we expect that each word $w^t$ in target language is generated either from a target event or a source event as follows:
$ p(w^t|\textbf{q}^s)= p(w^t|\textbf{z}^t)p(\textbf{z}^t|\textbf{q}^s) + p(w^t|\textbf{w}^s)p(\textbf{w}^s|\textbf{q}^s)$
in which $\textbf{z}^t$ is a topical variable on $F^t$ and $\textbf{w}^s$ is a translation of $w^t$ in the dictionary (see whole the details in~\cite{Ganguly:2012}).

It is noteworthy that topic modeling considers co-occurrences of the terms within documents without considering their sentence-level contexts.
Language modeling based on neural networks is a popular technique for capturing co-occurrences of the terms within a constant window $c$ and thus, this model embeds semantic of the language as well as deeper levels of n-grams~\cite{Mikolov:2013}.

But when it comes to cross-lingual environments a constraint is required for integrating dimensions in the languages. 
There are a couple of methods to this end; on-line methods and off-line methods. On-line methods aim at finding a unified space for the languages during the learning process as follows~\cite{Gouws:2014}: $\mathcal{L}(\theta) = \mathcal{L}(\theta^s) + \mathcal{L}(\theta^t) + \lambda f(\theta^s,\theta^t)$
where the last term is a regularization term.
Recently, Vulic et al., introduced the shuffling-based word embedding method over comparable corpora in which the alignments play as a number of constraints on the vectors~\cite{Vulic:2015b}. 
Ultimately, this approach considers neither the absolute/relative positions of the terms nor their sentence level contexts for estimation.

Off-line methods learn the vectors separately and then find a transformation matrix minimizing a constraint function $f$ as follows~\cite{Mikolov:2013b}:
$f(\textbf{W}) =  \sum_{\textbf{x},\textbf{z}}  || \textbf{W}^T\textbf{x} -\textbf{z}||^2$
in which $\textbf{x} \in \mathbb{R}^{n\times 1}$ and $\textbf{z} \in \mathbb{R}^{n\times 1}$ are the vectors of translation pairs from a dictionary or a parallel corpus.


\vspace{-0.3cm}
\section{Linear Projection between Languages based on Pseudo-relevant Documents}
\label{Linear Projection between Languages based on Pseudo comparable Documents}
\vspace{-0.3cm}
In this section we introduce the proposed method in more details.
We employ an off-line approach for learning bilingual representations of the words by exploiting pseudo-relevant documents in both source and target languages.
To this end, first we learn word representations of the pseudo-relevant collections separately and then focus on finding a transformation matrix minimizing a distance function between all translation pairs appeared on the collections.
As shown in Equation~\ref{eq:obj func} our goal is to minimize $f$ with respect to a transformation matrix $\textbf{W} \in \mathbb{R}^{n\times n}$; $f$ is defined as follow:

\begin{equation}
\label{eq:obj func}
    f(\textbf{W}) =\sum_{(w^s,w^t)}  \frac{1}{2}|| \textbf{W}^T\textbf{u}_{w^s} - \textbf{v}_{w^t}||^2
\end{equation}
where, $w^t$ is a translation pair of $w^s$ from $F^s$ that is appeared in $F^t$; $\textbf{u}_{w^s}\in \mathbb{R}^{n\times 1}$ and $\textbf{u}_{w^t} \in \mathbb{R}^{n\times 1}$ are the corresponding vectors respectively.
To solve this problem we choose the stochastic gradient descent algorithm (i.e., $\frac{\partial f}{\partial \textbf{W}}=0$):

\begin{equation}
\textbf{W}^{t+1} = \textbf{W}^{t} - \eta(\textbf{W}^T\textbf{u}_{w^s} - \textbf{v}_{w^t})\textbf{u}_{w^s}^T
\end{equation}

where  $\eta$ is a constant learning rate. $\textbf{W}$ is to be initialized randomly and then be updated incrementally.
\vspace{-0.3cm}
\subsection{Bilingual Representations and Translation Models}
In this section we introduce a method based on bilingual word representations for building a translation model for query and then incorporate \jComment{it} within language modeling, the state-of-the-art retrieval framework. 

\begin{equation}
\hat{\textbf{u}}_{w^s}=\textbf{W}^T\textbf{u}_{w^s}
\end{equation}
where $\textbf{W}^T\textbf{u}_{w^s}$ transforms the source vector built on the source collection to the target low-dimensional space.
The new translation model is built as follows:
\begin{equation}
p(w_t|w_s) = \frac{e^{\frac{\hat{\textbf{u}}_{w^s}.\textbf{v}_{w^t}}{||\hat{\textbf{u}}_{w^s}||~||\textbf{v}_{w^t}||}}}{\sum\limits_{\bar{w}^t\in \mathscr{T}\{w^s\}}\!\!\!\!\!\!\!e^{\frac{\hat{\textbf{u}}_{w^s}.\textbf{v}_{\bar{w}^t}}{||\hat{\textbf{u}}_{w^s}|| ~||\textbf{v}_{\bar{w}^t}||}}}
\end{equation}

where $\mathscr{T}\{w^s\}$ is the list of translations of $w^s$.
Instead of topical information propagation taking place on CLTRLM and joint cross-lingual topical relevance model (JCLTRLM), CLWETM tailors semantic projection and scaling both embedded in $\textbf{W}$.
In other words, CLWETM embeds quite sentence-level contexts of the words and thus captures deeper levels of n-grams as well.
\vspace{-0.4cm}
\subsubsection{Combining Translation Models}
Since \jComment{the obtained model} is a probabilistic translation model we can interpolate it with other models using a constant controlling parameter as follow:
\begin{equation}
\label{eq:interpolate}
   p(w^t|w^s) = \alpha p_1(w^t|w^s) + (1-\alpha) p_2(w^t|w^s)
\end{equation}
\vspace{-0.9cm}
\section{Experiments}
\vspace{-0.2cm}
\subsection{Experimental Setup}
\vspace{-0.2cm}
\begin{table*}[t]
\centering
\caption{Collection Characteristics}
\begin{tabular}{|c|c|c|c|c|c|c|} \hline
ID & Lang. & Collection & Queries & \#docs  & \#qrels \\ \hline
IT & Italy &  \specialcell{La Stampa 94, AGZ 94} & \specialcell{CLEF 2003-2003, Q:91-140} &  108,577& 4,327\\ \hline
SP & Spanish & EFE 1994 & CLEF 2002, Q:91-140 & 215,738 &  1,039\\ \hline
DE & German & \specialcell{Frankfurter Rundschau 94, \\SDA 94, Der Spiegel 94-95} & CLEF 2002-03, Q:91-140 & 225,371 &  1,938\\ \hline
FR & French & \specialcell{Le Monde 94, \\SDA French 94-95} & CLEF 2002-03, Q:251-350 & 129,806 & 3,524 \\ \hline
\end{tabular}
\label{tab:dataset}
\end{table*}

\newcommand{\thickhline}{%
    \noalign {\ifnum 0=`}\fi \hrule height 1pt
    \futurelet \reserved@a \@xhline
}

\begin{table*}[t]
\caption{Comparison of different dictionary-based short query translation methods. Superscripts indicate that the improvements are statistically significant (2-tail t-test, $p \leq 0.05$). $n-m$ indicates all methods in range $[n,..,m]$.}
\resizebox{0.9\columnwidth}{!}{%
\begin{tabularx}{\textwidth}{c|c|c|c|c||c|c|c||c|c|c||c|c|c|}
\cline{2-14}
      &  &   \multicolumn{3}{c||}{FR (short)}  & \multicolumn{3}{c||}{DE (short)} & \multicolumn{3}{c||}{ES (short)} & \multicolumn{3}{c|}{IT (short)} \\ \cline{2-14} 
    &ID & MAP & P@5 & P@10 & MAP & P@5 & P@10 & MAP & P@5 & P@10 & MAP & P@5 & P@10\\ \cline{1-14} 
\rowcolor{blue}
\multicolumn{1}{|l|}{-}&MONO  & 0.3262 & 0.4121 & 0.3737 & 0.2675 & 0.4323 & 0.3688 & 0.3518 & 0.4962 & 0.4321 & 0.2949 & 0.3677 & 0.3115 \\ \hdashline
\multicolumn{1}{|l|}{1}&MT &0.2858&	0.3939	&0.3394	&0.2889	&0.4375	&0.3896	&0.3339	&0.4280	&0.3800	&0.2579	&0.3510	&0.3224 \\ \cline{1-14}\cline{1-14}\cline{1-14}
\tabucline{1-14} 
\multicolumn{1}{|l|}{2}&TOP-1 & 0.2211 & 0.3122 & 0.2735 & 0.2015 & 0.2531 & 0.2327 & 0.2749 & 0.3673 & 0.3265 & 0.1566 & \textbf{0.2208} & 0.1896 \\ \cline{1-14} 
\multicolumn{1}{|l|}{3}&UNIF  & 0.1944 & 0.2694 & 0.2357 & 0.2148 & 0.2816 & 0.2367 & 0.2362 & 0.2939 & 0.2490  & 0.1526 & 0.2000    & 0.1562 \\ \cline{1-14} 
\multicolumn{1}{|l|}{4}&STRUCT& 0.1677 & 0.25   & 0.226  & 0.1492 & 0.2267 & 0.2044 & 0.2472 & 0.3348 & 0.3283 & 0.0994 & 0.1333 & 0.1178 \\ \cline{1-14} 
\multicolumn{1}{|l|}{5}&BiCTM & 0.2156 & 0.3143 & 0.2755 & 0.2126 & 0.2816 & 0.2612 & 0.2652$^{*}$ & 0.3429 & 0.3163 & 0.1504 & 0.2167 & 0.1771 \\
\cline{1-14} 
\multicolumn{1}{|l|}{6} &JCLTRLM & 0.1735 & 0.2687 & 0.2417 & 0.1416 & 0.2178 & 0.1933 & 0.2358 & 0.3522 & 0.3283 & 0.1105 & 0.1733 & 0.1511 \\
\cline{1-14}
\multicolumn{1}{|l|}{7} & MIXWETM & 0.2202 & 0.3143 & 0.2622 & \textbf{0.2166} & \textbf{0.2166} & \textbf{0.2633} & 0.2790  & 0.3755 & 0.3122 & 0.1587 & 0.2125 & 0.1833\\
\cline{1-14}
\multicolumn{1}{|l|}{8}&\small{CLWETM}& \textbf{0.2312}$^{2-7}$ & \textbf{0.3306} & \textbf{0.2806} & 0.2158$^{246}$ & 0.2816 & 0.2551 & \textbf{0.2915}$^{2-7}$ & \textbf{0.3837} & \textbf{0.3367} & \textbf{0.163}$^{2-7}$  & \textbf{0.2208} & \textbf{0.1937} \\  \cline{1-14} 
\end{tabularx}}
\label{tab:exp2}
\end{table*}
\begin{table*}[t]
\caption{Comparison of different dictionary-based long query translation methods. }
\resizebox{0.89\columnwidth}{!}{%
\begin{tabularx}{\textwidth}{c|c|c|c|c||c|c|c||c|c|c||c|c|c|}
\cline{2-14}
    &    &   \multicolumn{3}{c||}{FR (long)}  & \multicolumn{3}{c||}{DE (long)} & \multicolumn{3}{c||}{ES (long)} & \multicolumn{3}{c|}{IT (long)} \\ \cline{2-14}
    &ID & MAP & P@5 & P@10 & MAP & P@5 & P@10 & MAP & P@5 & P@10 & MAP & P@5 & P@10\\ \cline{1-14} 
\multicolumn{1}{|l|}{-}&MONO  & 0.4193 & 0.5354 & 0.4727 & 0.3938 & 0.5280 & 0.4780 & 0.5281 & 0.6720 & 0.5960 & 0.3947 & 0.5022 & 0.4356 \\
\hdashline
\multicolumn{1}{|l|}{1}&MT  &0.3395 & 0.4263 & 0.3747 & 0.3436 & 0.4400  & 0.4280 & 0.4208 & 0.5600  & 0.478 & 0.1376 & 0.1551 & 0.1306\\
\cline{1-14} \cline{1-14} \cline{1-14} 
\tabucline{1-14}
\multicolumn{1}{|l|}{2}&TOP-1 & 0.3077 & 0.3960  & 0.3434 & 0.2242 & 0.3080 & 0.2500  & 0.3762 & 0.4800  & 0.4320 & 0.2195 & 0.2800   & 0.2622 \\ \cline{1-14} 
\multicolumn{1}{|l|}{3}&UNIF  & 0.2709 & 0.3556 & 0.3091 & 0.2425 & 0.2840 & 0.2540 & 0.3243 & 0.3680 & 0.3340 & 0.2095 & 0.2311 & 0.2000    \\ \cline{1-14} 
\multicolumn{1}{|l|}{4}&STRUCT& 0.1800   & 0.2646 & 0.2394 & 0.2103 & 0.252 & 0.2500  & 0.2951 & 0.4000   & 0.3760 & 0.1942 & 0.2444 & 0.2244 \\ \cline{1-14} 
\multicolumn{1}{|l|}{5}&BiCTM & 0.3050  & 0.3899 & 0.3505 & 0.2442 & 0.3280 & 0.2780 & 0.3841 & 0.4640 & 0.4340 & 0.2172 & 0.2622 & 0.2422 \\ \cline{1-14} 
\multicolumn{1}{|l|}{6}&JCLTRLM & 0.2266 & 0.3414 & 0.2990  & 0.1520  & 0.2160 & 0.1880 & 0.2734 & 0.404 & 0.3500  & 0.1459 & 0.2133 & 0.1756 \\ \cline{1-14}
\multicolumn{1}{|l|}{7}&MIXWETM & 0.2983 & 0.3919 & 0.3485 & \textbf{0.2652} & 0.3400  & 0.3040 & 0.3677 & 0.4280 & 0.4080 & \textbf{0.2381} & \textbf{0.3022} & \textbf{0.2733} \\ \cline{1-14}
\multicolumn{1}{|l|}{8}&\small{CLWETM}& \textbf{0.3167}$^{2-7}$ & \textbf{0.4101} & \textbf{0.3657} & 0.2622$^{2-6}$ & \textbf{0.3480} & \textbf{0.3080} & \textbf{0.4029}$^{2-7}$ & \textbf{0.500}   & \textbf{0.4620} & 0.2380$^{2-6}$  & 0.2978 & 0.2667 \\ \cline{1-14} 
\end{tabularx}
}
\label{tab:exp3}
\end{table*}

Overview of the used collections is provided in Table~\ref{tab:dataset}. 
The source collection is a pool of Associated Press 1988-89, Los Angeles Times 1994, and Glasgow Herald 1995 collections that are used in previous TREC and CLEF evaluation campaigns for ad-hoc retrieval.

In all experiments, we use the language modeling framework with the KL-divergence retrieval model and Dirichlet smoothing method to estimate the document language models, where we set the dirichlet prior smoothing parameter $\mu$ to the typical value of $1000$.
To improve the retrieval performance, we use the mixture model for pseudo-relevance feedback with the feedback coefficient of $0.5$. 
The number of feedback documents and feedback terms are set to the typical values of $10$ and $50$, respectively.

All European dictionaries, documents, and queries are normalized and stemmed using the Porter stemmer.  
Stopword removal is also performed.\footnote{We use the stopword lists and the normalizing techniques available at \url{http://members.unine.ch/jacques.savoy/clef/}.}
The Lemur toolkit\footnote{\url{http://www.lemurproject.org/}} is employed as the retrieval engine in our experiments.

We use the Google dictionaries in our experiments\footnote{\url{http://translate.google.com}}. 
In the European languages, we do not transliterate out of vocabulary (OOV) terms of the source languages. 
The OOVs of the target language are used as their original forms in the source documents, since they are cognate languages. 
Note that we use the uniform distribution approach as the initial translation model for retrieving top documents. 
It is worth mentioning that $p(w^t|\textbf{w}^s)$ in CLTRLM (see Sec. \ref{sec:related works}) is estimated by a bi-gram coherence translation model (BiCTM) introduced in~\cite{Monz:2005}. 
Weights of the edges of the graph are estimated by $p(w_j|w_i)$ computed by SRILM toolkit~\footnote{http://www.speech.sri.com/projects/srilm/}.
BiCTM is also used as $p_2$ where $\alpha$ is set by 2-fold cross-validation (see Equation~\ref{eq:interpolate}).

As discussed in Section~\ref{Linear Projection between Languages based on Pseudo comparable Documents}, we used stochastic gradient descent for learning $\textbf{W}$ which is initialized with random values in $[-1,1]$; 
$\eta$ is set to a small value which also decreases after each iteration. $\textbf{u}_{w^s}$ and $\textbf{v}_{w^t}$ are computed based on negative sampling skip-gram introduced in~\cite{Mikolov:2013}; the size of the window, the number of negative samples, and the size of the vectors are set to typical values of $10$, $45$, and $50$ respectively.

As shown in~\cite{Ganguly:2012} JCLTRLM outperforms CLTRLM and therefore we opted JCLTRLM as a baseline. 
The parameters of LDA are set to the typical values $\alpha_d=0.5$ and $\beta_d=0.01$.
Number of topics in JCLTRLM is obtained by 2-fold cross-validation.
\vspace{-0.4cm}
\subsection{Performance Comparison and Discussion}
\vspace{-0.2cm}
In this section we want to compare \jComment{effectiveness} of a number of competitive methods in CLIR.
We consider the following dictionary-based CLIR methods to evaluate the proposed method: (\textit{1}) the top-1 translation of each term in the bilingual dictionaries (TOP-1), (\textit{2}) all the possible translations of each term with equal weights (UNIFORM), (\textit{3}) (BiCTM) proposed in~\cite{Monz:2005} , (\textit{4}) the JCLTRLM method proposed in~\cite{Ganguly:2012}, and (\textit{5}) MIXWETM \cite{Vulic:2015b}. 
As bases of comparisons we also provided results of the monolingual runs in each collection (MONO) and Google machine translator (MT).
However, our main focus is to investigate superiority of the proposed method compared to the dictionary-based CLIR baselines which are available for almost all pairs of languages.

\begin{figure*}[t]
        \vspace{-0.5cm}
        \centering
        \begin{subfigure}[b]{0.239\textwidth}
                \includegraphics[width=\textwidth]{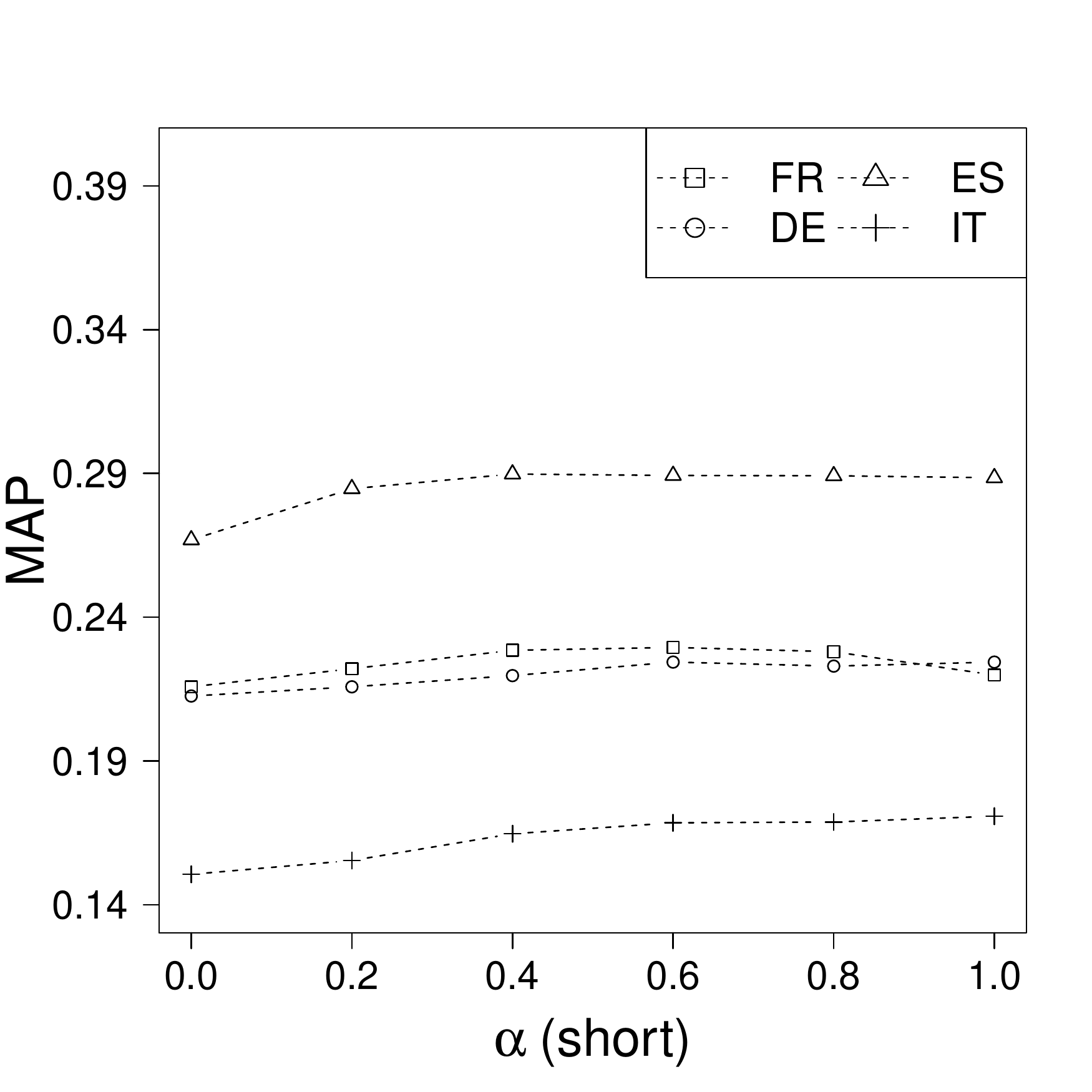}
        \end{subfigure}%
        ~
        \begin{subfigure}[b]{0.239\textwidth}
            \includegraphics[width=\textwidth]{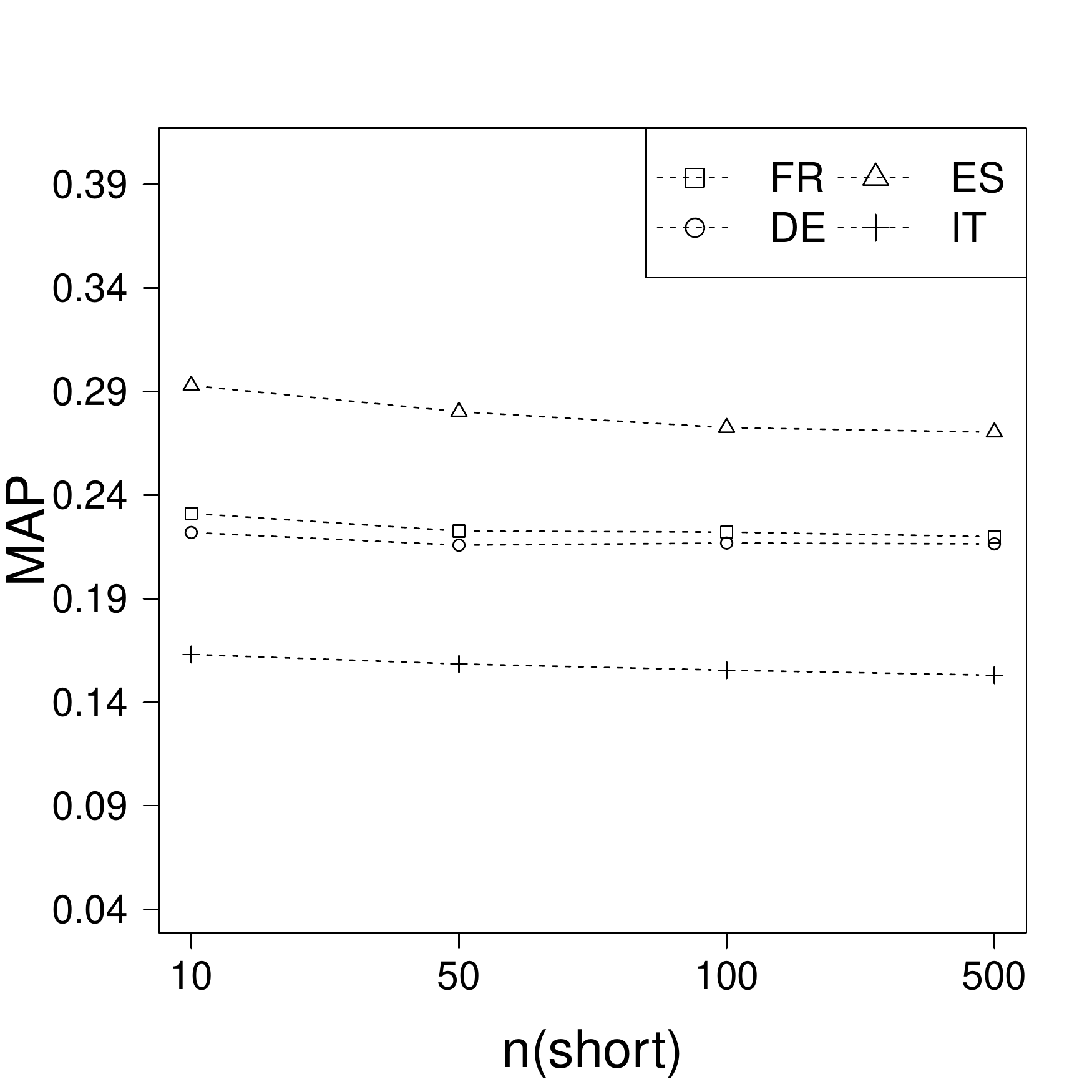}
        \end{subfigure}
        ~
        \begin{subfigure}[b]{0.239\textwidth}
                \includegraphics[width=\textwidth]{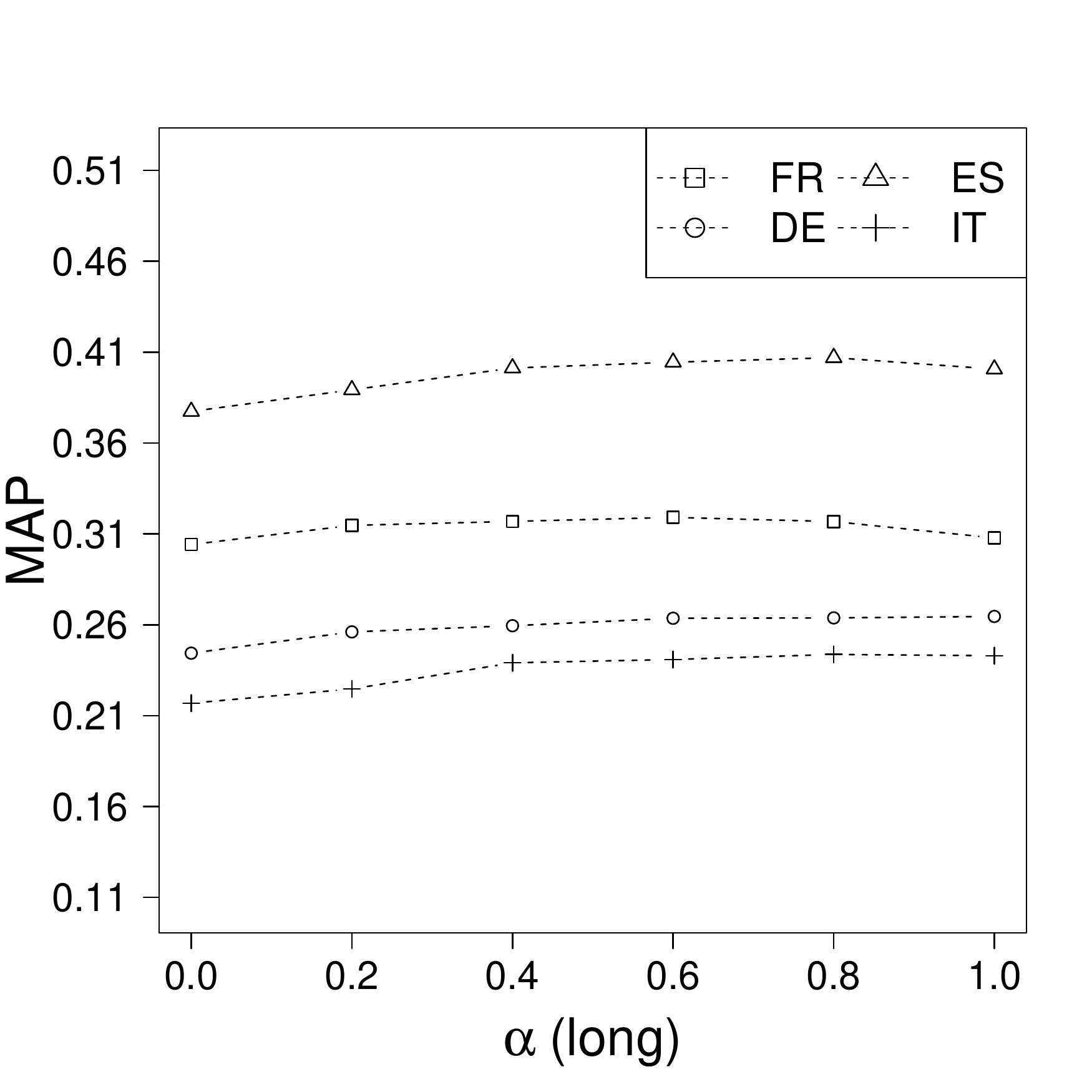}
        \end{subfigure}%
        ~
        \begin{subfigure}[b]{0.239\textwidth}
            \includegraphics[width=\textwidth]{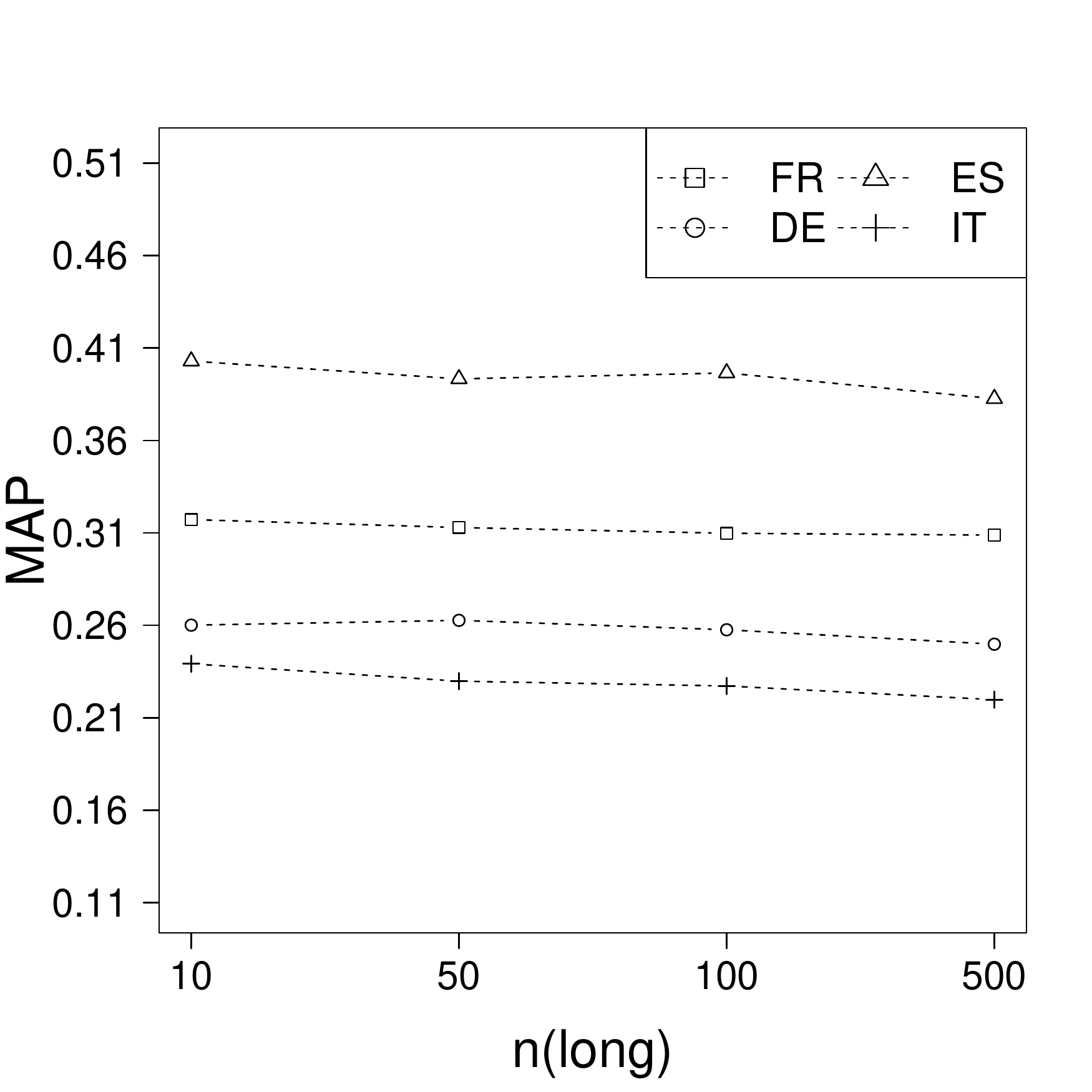}
        \end{subfigure}
        ~ 
        \vspace{-0.3cm}
        \caption{MAP sensitivity of CLWETM to $\alpha$ and the number of feedback documents.}
        \vspace{-0.3cm}
        \label{fig:param1}
\end{figure*}

All the results (on short queries) are summarized in Table~\ref{tab:exp2} and Fig. \ref{fig:param1} shows sensitivity of CLWETM to $\alpha$ and the number of feedback documents.
As shown in the table, both MIXWETM and CLWETM outperform other methods in terms of MAP, P@5, and P@10 in all the collections. 
MIXWETM and CLWETM consistently achieved better results compared to others, but CLWETM is clearly more effective than MIXWETM in almost all the datasets.
Although CLWETM lost the competition to MIXWETM in DE, but the differences are not statistically significant.
One reason for this outcome is the lower performance of BiCTM compared to other collections (see Eq. \ref{eq:interpolate}, Table \ref{tab:exp2}, and Fig. \ref{fig:param1}). 
Another reason can be the lower sensitivity of the method to $n$ in this collection.
As shown in Fig. \ref{fig:param1} top-ranked documents in DE are not as helpful as FR, ES, and IT and thus neither CLWETM nor MIXWETM has significant improvements compared to BiCTM.  

Although the focus of this research is on dictionary-based CLIR, but it is clear to a reader that in the European collections with short queries the results of MT are higher than all the baselines in dictionary-based CLIR.
In the rest of the experiments, we shed light on effectiveness of the methods on verbose queries obtained by concatenating title and description parts of the topics.
Table \ref{tab:exp3} shows the results; the results also confirm the effectiveness of CLWETM over the dictionary-based methods.
The most interesting point is decrements of the gaps between CLWETM and MT in quite all the collections. 
CLWETM reached 93.2\%, 76.3\%, 95.7\%, and 172.9\% of the performance of MT in terms of MAP in ES, DE, ES, and IT respectively.
In IT we see noticeable decrement of the performance by MT on the verbose queries where the dictionary-based techniques are quite stable.

\vspace{-0.4cm}
\subsection{Parameter Sensitivity}
\vspace{-0.2cm}
We investigate the sensitivity of the proposed method to two parameters $\alpha$
and $n$ in Figure \ref{fig:param1}. 
We first fix one parameter to its optimal value and then try to get optimal value of the other one .
It demonstrates that both parameters work stably across FR, ES and IT collections. 
The optimal $\alpha$ value empirically is $0.6$ and the optimal value of $n$ is $10$ in almost all the collections. 

\vspace{-0.4cm}
\section{Conclusion and Future Works}
\vspace{-0.2cm}
In this paper we presented a translation model for cross-lingual information retrieval, that uses feedback documents in source and target languages for creating word vectors in each one, and then learns a projection matrix to project word vectors in the source language to their translations in the target language.
Then we introduced a method for building a translation model that can be easily interpolated with other models using a constant controlling parameter. 
We investigated the performance of the proposed method on four European collections of CLEF.
Our method showed more improvements in FR, SP, and IT since top-ranked documents in DE are not as comparable as the previous ones; therefor applying the proposed method on comparable corpora is considered as an interesting future work.
The proposed method reaches up to 87\% performance of machine translation (MT) in short queries and considerable improvements in verbose queries.
\vspace{-0.4cm}
\bibliographystyle{splncs03.bst}
\scriptsize{\bibliography{sigproc.bib}}
\end{document}